\documentclass[12pt,preprint]{aastex}

\shorttitle{AO Imaging of Nearby Young Stars}
\shortauthors{Brandeker, Jayawardhana \& Najita}

\begin{document}

\title{Keck Adaptive Optics Imaging of Nearby Young Stars: \\ Detection of
Close Multiple Systems}

\author{Alexis Brandeker}
\affil{Stockholm Observatory, AlbaNova University Center,
 SE-106 91 Stockholm, Sweden}
\email{alexis@astro.su.se}

\author{Ray Jayawardhana}
\affil{Department of Astronomy, University of Michigan, Ann Arbor, MI 48109, U.S.A.}
\email{rayjay@umich.edu}

\and 
\author{Joan Najita}
\affil{National Optical Astronomy Observatory, Tucson, AZ 85719, U.S.A.} 
\email{najita@noao.edu}

\begin{abstract}
Using adaptive optics on the Keck II 10-meter telescope on Mauna Kea, we 
have surveyed 24 of the nearest young stars known in search of close 
companions. Our sample includes members of the MBM\,12 and TW\,Hydrae 
young associations and the classical T\,Tauri binary UY\,Aurigae in the 
Taurus star-forming region. We present relative photometry and accurate
astrometry for 10 close multiple systems. The multiplicity frequency 
in the TW\,Hydrae and MBM\,12 groups are high in comparison to other young 
regions, though the significance of this result is low because of 
the small number statistics. We resolve S\,18 into a triple system including
a tight 63\,mas (projected separation of 17\,AU at a distance of 275\,pc) 
binary for the first time, with a hierarchical configuration reminiscent of 
VW\,Chamaeleontis and T\,Tauri. Another tight binary in our sample -- 
TWA\,5~Aab (54\,mas or 3\,AU at 55\,pc) -- offers the prospect of 
dynamical mass measurement using 
astrometric observations within a few years, and thus could 
be important for testing pre-main sequence evolutionary models. Our 
observations confirm with 9$\sigma$ confidence that the brown dwarf 
TWA\,5~B is bound to TWA\,5~A. We find that the flux ratio of UY\,Aur 
has changed dramatically, by more than a magnitude in the $H$-band, 
possibly as a result of variable extinction. With a smaller flux ratio, 
the system may once again become detectable as an optical binary, as it 
was at the time of its discovery in 1944. 
Taken together, our results demonstrate that adaptive optics on large
telescopes is a powerful tool for detecting tight companions, and thus
exploring the frequency and configurations of close multiple systems. 

\end{abstract}

\keywords{binaries: close -- circumstellar matter --
stars: pre-main-sequence --
techniques: high angular resolution}

\section{Introduction}
Nearby young stellar associations offer unique advantages for detailed studies
of star and planet formation (Jayawardhana \& Greene 2001). In particular, their 
proximity in combination
with modern adaptive optics facilitates sensitive studies of individual
star systems down to physical scales of $\sim$3\,AU, closing the gap between
spectroscopic and visual binaries and thereby offering the prospect of 
a complete census of multiplicity among young pre-main sequence (PMS) stars 
down to sub-stellar masses. Completeness in sampling the binary frequency is
important to avoid bias in the understanding of formation and evolution
of multiple star systems. Several studies have shown multiplicity frequencies 
to be significantly higher among PMS stars than their main sequence (MS) 
counterparts (Duch\^{e}ne 1999). Suggested explanations for this discrepancy 
include observational sample/sensitivity bias, environmental or evolutionary 
effects, such as disruption of binaries by close approaches of other stars 
(Ghez, Neugebauer, \& Matthews 1993), ejections of low-mass companions in 
multiple systems (e.g., Bate et al. 2002), or even stellar mergers. 

Historically, the determination of MS binary orbits, and thus their 
dynamical masses, have been critical for the successful theory 
of stellar structure developed in the early 20th century. Unfortunately, 
the nearest known PMS stars, at a distance of $\sim$150 pc, had been 
until recently too far away to resolve companions with reasonably short 
orbital periods. Resolving companions spatially is important since 
spectroscopic observations by themselves only yield the relative masses
of binaries.
By resolving a spectroscopic binary spatially down to a physical
scale of a few AU, it is not only possible to determine the dynamical masses
of the components, but also to derive an independent distance to the system. 
Therefore, spatially resolved PMS binaries are essential tools for 
testing models of PMS stellar structure and evolution (Palla \& Stahler 
2001). 
An interesting recent example is the dynamical mass determination of
the T\,Tauri binary NTT\,045251+3016 by Steffen et al. (2001);
using precise radial velocity and astrometric data, they show that it 
is now possible to derive the stellar masses without any astrophysical 
assumption and compare to PMS evolutionary tracks. 

A third important reason to search for companions of PMS stars is to 
clear up the ambiguity when placing them on evolutionary luminosity-color 
diagrams. An unresolved PMS binary will show a luminosity (and possibly 
color) different from the individual components, thereby biasing any 
conclusions drawn from evolutionary diagrams (for an extensive review, 
see Mathieu 1994). 

Due to their proximity, the nearby young associations MBM\,12 
(Luhman 2001) and TW\,Hydrae (Kastner et al. 1997) are  prime targets for 
detailed studies of circumstellar disks and close companions. Jayawardhana
et al. (1999b) conducted a census of disks among TW\,Hydrae association 
(TWA) members and Jayawardhana et al. (1999a) showed that most of the mid-infrared 
emitting dust in the 1\farcs4 binary Hen\,3-600 is associated with the 
primary. A similar study of MBM\,12 members was reported by Jayawardhana 
et al. (2001), who found that six of the eight stars observed harbored 
significant mid-infrared excess; four of these disks have now also been 
detected in the (sub-)millimeter (Itoh et al. 2003; Hogerheijde et al. 
2003). Using adaptive optics (AO) on Gemini North, Jayawardhana et al. 
(2002) resolved LkH$\alpha$ 263 (MBM\,12-3) into a 0\farcs415 binary 
and imaged a companion edge-on disk 4\farcs115 away. Chauvin et al. 
(2002) reported several additional binaries in MBM\,12, and 
Macintosh et al. (2001) reported a 55~mas binary in TWA. In this work,
we report the results of an AO survey of the inner 
(sub-arcsecond) regions of UY\,Aur and member stars of MBM\,12 and TWA, 
in search of close companions. Among the 24 stars we observed, we found 
six close binaries and a triple not known at the time, and confirmed four 
known binaries.

\section{Observations}
We surveyed 24 young stars on 2000 February 22--23 with the 10-m Keck~II
telescope on Mauna Kea, Hawaii, using adaptive optics (Wizinowich et al. 
2000) and the near-infrared (NIR) camera KCam. KCam is based on a 
256$\times$256 pixel NICMOS-3 HgCdTe array, and was provided by the University
of California as an early interim engineering grade camera. The pixel
scale was measured on 2000 February 21 by Macintosh et al. (2001) to be
$17.47 \pm 0.06$~mas\,pixel$^{-1}$. Since one 128$\times$128 quadrant of
KCam is non-functional, the field of view was L-shaped with 2\farcs24
per quadrant. We observed the targets by co-adding 1 to 20 exposures
of integration time 0.62\,s to 60\,s each, and then moving the target
to another quadrant of the array so that the target was located in each
of the three working quadrants at least once. The science target itself 
was used as a wave front sensor for the AO system, except in the case of
TWA\,9~B, where we used TWA\,9~A. We observed through three different filters,
$J$ (1.26\,\micron), $H$ (1.648\,\micron) and $K^{\prime}$ (2.127\,\micron),
located in a cold filter wheel. The Strehl ratio during the two nights was
0.20--0.25 in $H$.

\section{Data reduction and analysis}
The data reduction was done in a standard way by subtracting sky
frames from source frames and then dividing by a flat obtained on-sky
during dusk. We also corrected for the substantial number of bad pixels.
Cosmic ray hits were less of a problem because the exposures were short.
For a frame with the target in a given quadrant of the array,
the sky frame was obtained by averaging the frames with the target in
one of the other two quadrants. For binaries with separations greater
than $\sim$2\arcsec, the secondary was only visible on one frame out of
three, and only one frame was free from sources in the right quadrants
to be used as a sky.

Due to the small field of view and the variability of the AO point spread
function (PSF), it is not possible to obtain accurate absolute photometry. 
Relative photometry and astrometry of binary systems were measured on
multiple frames, and then combined to get an estimate of the errors. We
did aperture photometry with a diameter of 4 pixels, except for the
close binaries in the TWA\,5~A and S\,18~B systems where we used apertures 
with a diameter of only 2 pixels. The contamination from the
companion into the aperture was estimated by placing an aperture also on
the opposite side of the companion. To obtain accurate astrometry
of these tight binaries we used the myopic deconvolution algorithm IDAC 
(Christou et al. 1999). IDAC estimates the PSF simultaneously with the 
deconvolved image, but needs a series of exposures with varying PSFs 
and/or accurate estimates of the PSF. For TWA\,5~A we used 7 frames 
in $H$, but for S\,18~B we had only 3 $H$-frames. Since S\,18 is a 
triple, however, we could make use of the primary S\,18~A as an 
accurate simultaneous PSF estimate. Based on comparisons with Jayawardhana
et al. (2002) and Chauvin et al. (2002), the systematic errors in our 
astrometry due to array orientation uncertainties are likely to be 
$\sim$ $1\fdg$

\begin{figure}
\plotone{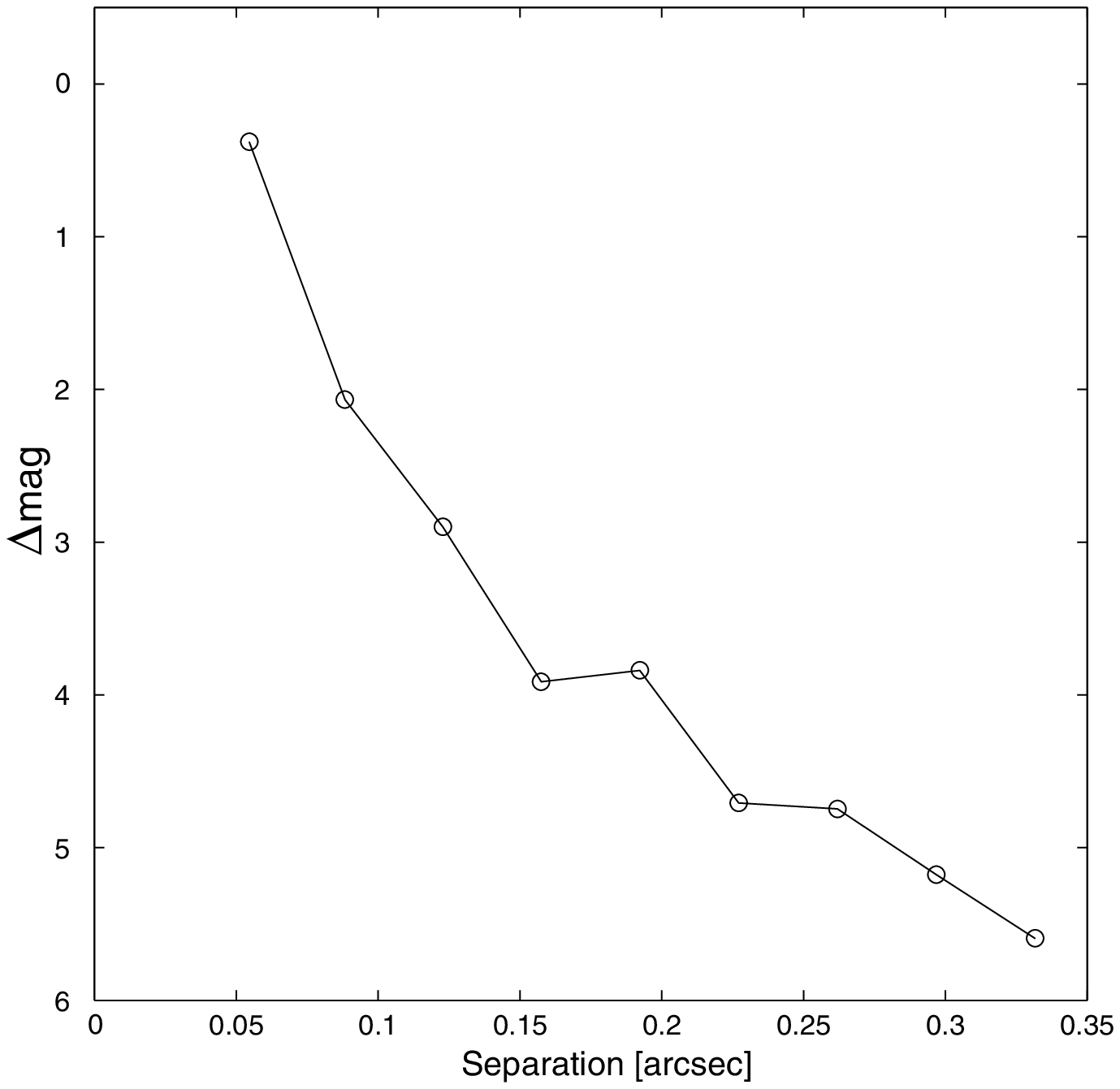}
\caption{This is the approximate 5$\sigma$ detection contrast sensitivity as a
function of separation. The inner 0\farcs3 are dominated by speckle noise, while
sky noise causes the curve to flatten out to a flux ratio of $\Delta {\rm mag}=$\,6--7.
\label{fig-1}}
\end{figure}

In order to estimate our sensitivity to finding close faint companions, 
we measured the speckle noise from observed PSFs of single stars. Our
approximate 5$\sigma$ detection limit as a function of separation
is shown in Fig.~\ref{fig-1}.

\section{Results and discussion}

Among the 9 proposed members in the MBM\,12 association we observed, we
found 5 to be binaries, all previously reported by Chauvin et al. (2002),
and one to be a new triple (Table~\ref{tbl-1}). To calculate the multiplicity
frequency of the 12 proposed members of MBM\,12 (Luhman 2001), we 
assume that LkH$\alpha$\,262 and LkH$\alpha$\,263 are part of the same 
quadruple system (Jayawardhana et al. 2002), and count 
LkH$\alpha$\,264 and MBM\,12-10 as binaries (Chauvin et al. 2002) 
\footnote[1]{HD\,17332, RXJ\,0255.3+1915 and RXJ\,0306.1+1921 are not counted
because they are not likely to be PMS members of MBM\,12; see Jayawardhana
et al. (2001) and Luhman (2001).}. 
The multiplicity frequency, defined as $mf=\frac{b+t+q}{s+b+t+q}$ where 
$s$, $b$, $t$ and $q$ are the number of single, double, triple and 
quadruple systems respectively, is found to be 
$\frac{5+1+1}{4+5+1+1} = 0.64\pm0.16$, where the quoted error is
statistical. Similarly, the average number of 
companions per star system, defined as $csf=\frac{b+2t+3q}{s+b+t+q}$
(Duch\^{e}ne 1999), equals 
$\frac{5+2\times1+3\times1}{4+5+1+1} = 0.91\pm0.30$. 

In TWA we found no new companion candidates apart from the TWA\,5~A 
54\,mas binary, which was also found independently by Macintosh 
et al. (2001). The corresponding multiplicity numbers for TWA are 
$mf=0.58\pm0.12$ and $csf=0.84\pm0.22$, where we have counted the
companions in the systems TWA\,1--19, as reported by Webb et al. (1999) 
and Zuckerman et al. (2001). TWA\,5 was assumed to be a quadruple system,
including the brown dwarf TWA\,5~B and a spectroscopic binary in
TWA\,5~A (see \ref{TWA5}). This only affects $csf$, as $mf$ is insensitive to
multiple companions. Both TWA and MBM\,12 have multiplicity numbers 
that are on the high side compared to those reported for other young 
associations (Duch\^{e}ne 1999), though the significance of this result 
is low due to the small number statistics.

In Table~\ref{tbl-2} we give the coordinates and absolute NIR fluxes 
for the components as inferred from 2MASS. We plainly divided the 
unresolved 2MASS flux to the stars according to our measured flux ratios. 
Note that the epoch of the 2MASS observations differ by $\sim$2~years 
from ours, so variations in flux may introduce errors in addition to the 
propagated errors quoted in the table. In Table~\ref{tbl-3} we present 
stars that did not reveal close companions in our survey. 

\subsection{S\,18}
\begin{figure}
\plotone{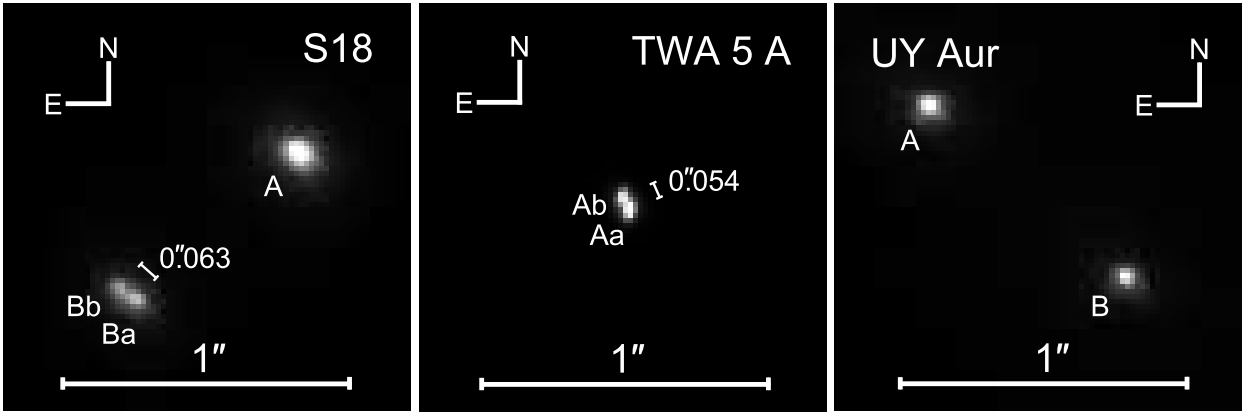}
\caption{These are the sharpest $H$-band frames we obtained of the young multiple star
systems S\,18, TWA\,5~A and UY\,Aur. The image scale is 17.5\,mas\,pixel$^{-1}$.
North is up and east is to the left.
\label{fig-2}}
\end{figure}

S\,18 in MBM\,12 was reported to be a 0\farcs753 binary by Chauvin et al. 
(2002). In our $H$-band observations, we resolve the secondary B into a 
tight Ba/Bb 63\,mas binary in itself (Fig.~\ref{fig-2}). Note the close 
similarity in configuration and scale with the triple stars VW\,Chamaeleontis  
(Brandeker et al. 2001) and T\,Tauri (Duch\^{e}ne, Ghez \& McCabe 2002). 
In $J$ and $K$ we were unable to resolve B sufficiently to measure the 
flux ratio accurately. If we assume the distance to MBM\,12 to be 
$\sim$275\,pc (Luhman 2001), then the projected distance between the
stars is $\sim$17\,AU. Using Kepler's third law and assuming a circular
orbit, we get the orbital period $P = 71\,(M_B\,{\alpha}^3)^{-1/2}~\rm{yr}$,
where $\alpha$ is the projection factor
$\alpha$ = observed (projected) separation / real separation,
and $M_B$ is the system mass of S\,18~Ba/Bb in units of $M_{\sun}$. This
corresponds to an average motion in position angle of 
$5.0\,(M_B\,{\alpha}^3)^{1/2}~\rm{deg\,yr^{-1}}$, and should
be readily detected within a year. Similarly, the relative radial 
velocity amplitude is $7.3\,(M_B\,{\alpha})^{1/2}{\rm sin}\,i~\rm{km\,s^{-1}}$, 
which may also be measurable over the period of a few decades unless 
the inclination $i$, the angle between the orbital plane and the plane
of sky, is very small.

\subsection{TWA\,5~A~\&~B\label{TWA5}}
We found TWA\,5~A to be a tight 54\,mas binary, as also reported by 
Macintosh et al. (2001). At the distance of TWA, ($\sim$55\,pc, Perryman 
et al. 1997), this corresponds to a projected separation of only 3\,AU. 
The orbital period is thus $P = 5.2\,(M_A\,{\alpha}^3)^{-1/2}~\rm{yr}$, 
corresponding to an average position angle motion of $69\,(M_A\,{\alpha}^3)
^{1/2}~\rm{deg\,yr^{-1}}$. The radial relative velocity amplitude is 
$17.3\,(M_A\,{\alpha})^{1/2}{\rm sin}\,i~\rm{km\,s^{-1}}$. Follow-up
studies of this system have the potential to obtain an accurate dynamical 
mass estimate within a few years. One possible complication may be the 
presence of an additional {\em spectroscopic binary} in the TWA\,5~A system, 
as suggested by Webb et al. (1999) and Torres et al. (2003). 

Our observations also reveal the brown dwarf companion TWA\,5~B discovered by
Lowrance et al. (1999). With a flux ratio of $\sim$100 (5 mag) and a separation 
of 1\farcs954, the companion was not always ideally placed in our L-shaped array. 
In our $K^{\prime}$-band observations the star was close to the edge, which accounts 
for large estimated errors in the photometry. The astrometry, however, is excellent,
and we can use this to increase the significance of the companion hypothesis, as
previously argued by Neuh\"{a}user et al. (2000). Weintraub et al. (2000) used the 
Hubble Space Telescope to measure accurate relative astrometry on 1998 July 12. 
By combining their astrometry with our measurements we conclude that the relative 
positions between TWA\,5~A and TWA\,5~B has changed by 
$\Delta\alpha=9.5\pm27\,\rm{mas}$ and $\Delta\delta=-6\pm10\,\rm{mas}$
during the 1.61\,yr between the epochs. The proper motion of TWA\,5~A, as found in the 
Tycho-2 catalog (H{\o}g et al. 2000), is $\mu_{\alpha}=-81.6\pm2.5\,\rm{mas\,yr^{-1}}$,
$\mu_{\delta}=-29.4\pm2.4\,\rm{mas\,yr^{-1}}$. The expected position of a background 
object is thus excluded at $9\sigma$, somewhat higher than the $5\sigma$ obtained by 
Neuh\"{a}user et al. (2000), and consistent with their estimated orbital motion 
$13.4\pm4.2\,\rm{mas\,yr^{-1}}$ of a bound system.

\subsection{UY\,Aur\label{UYAur}}

In addition to the young stars in the nearby associations MBM\,12 and TWA, we
also observed the classical T\,Tauri binary system UY\,Aur in the Taurus 
star-forming region. UY\,Aur is unique in that it was first discovered as an 
optical binary in 1944 by Joy \& van Biesbroeck (1944) with a flux ratio 
0.4--0.5 mag in $V$. By 1992, the companion was no longer visible in optical 
wavelengths and at least 5 mag fainter than the primary in $R$ (0.71\,\micron) 
(Herbst, Koresko, \& Leinert 1995). The system is obviously variable, with 
$K$-band flux ratios having varied from $\Delta K=0.76\pm0.06$\,mag (White 
\& Ghez 2001) to $\Delta K=1.38\pm0.08$\,mag (Leinert et al. 1993). The system 
was also the second detected to show a circumbinary disk (Dutrey et al. 1996; 
Duvert et al. 1998; Close et al. 1998). Our $H$-band observation of
UY\,Aur has a too small a field of view and is not sensitive enough to detect the
circumbinary disk. We find, however, that the flux ratio in $H$ is significantly
smaller, $1.47\pm0.01$, than previous measurements, $4.33\pm0.36$ (Close et al. 
1998). Unfortunately we can not assess whether it is the primary that has dimmed or
the secondary that has brightened (or both). Close et al. (1998) derived an extinction
of $A_V=1.0$\,mag and $A_V=9.2$\,mag for the primary and secondary respectively,
explaining the very red color of the secondary.
If the observed change in flux ratio in $H$ is due to a variable extinction, then the
implied change in extinction is $\Delta A_H=1.2$\,mag, corresponding to
$\Delta A_V=5.8$\,mag (assuming the extinction law $R_V=5.0$ in Table 1 of
Mathis 1990). A crude estimate,
using the spectral classes K7 and M0 for the primary and secondary respectively
(Herbst, Koresko \& Leinert 1995), suggests that the flux ratio in $R$ has evolved
from $\Delta R=6.6$ in 1996 October 24 (Close et al. 1995) to $\Delta R=2.0$ in
2000 February 22 (this work). The system may thus once again have become 
detectable as an optical binary, just as it was at the time of its discovery 
in 1944.

\section{Conclusions}
We have conducted a survey of the inner regions of nearby young star systems and
measured accurate astrometry of epoch 2000 February 22--23 for 8 binaries
and 2 triples, as well as flux ratios. Our main conclusions are: 

\begin{itemize}
\item[$\bullet$] The multiplicity frequency and average companion number of the
young associations MBM\,12 and TWA are very high, though the significance of this 
result is low because of small number statistics.

\item[$\bullet$] The T\,Tauri star S\,18 is actually a triple system,
with a tight binary (0\farcs063 $\sim$17\,AU). Its hierarchical configuration is
very similar to VW\,Cha and T\,Tau. Follow-up astrometric observations may constrain
the mass within a few decades.

\item[$\bullet$] The T\,Tauri star TWA\,5~A is resolved into a very tight binary
in itself (0\farcs054 $\sim$3\,AU). Follow-up astrometric and spectroscopic
observations within a few years have the potential to accurately measure the
dynamical mass of individual components as well as an independent distance,
thus providing an ideal system to test evolutionary models of PMS stars.

\item[$\bullet$] The brown dwarf TWA\,5~B is shown to be bound to TWA\,5~A
with 9$\sigma$ confidence.

\item[$\bullet$] The classical T\,Tauri binary star UY\,Aur has changed its flux
ratio dramatically by more than a magnitude in $H$. The secondary may have 
once again turned into an optically detectable companion, as it was when first discovered
in 1944.
\end{itemize}

\acknowledgments
We thank an anonymous referee for helpful comments. 
We would like to acknowledge the great cultural significance of Mauna Kea 
for native Hawaiians, and express our gratitude for permission to observe 
from its summit. We also thank the Keck Observatory staff for their 
outstanding assistance over the past several years. We have made use 
of the NASA/IPAC Infrared Science Archive, NASA's Astrophysics Data System 
Bibliographic Services, the SIMBAD database, and data products from the 
Two Micron All Sky Survey in our research. This work was supported in part 
by NASA Origins grant NAG5-11905 to R.J.


\clearpage

\clearpage

\clearpage
\begin{deluxetable}{lccccc}
\tabletypesize{\scriptsize}
\tablecaption{Multiple stars KCam AO observations 2000 February 22--23\label{tbl-1}}
\tablewidth{0pt}
\tablecolumns{6}
\tablehead{
\colhead{Object} &
\colhead{} &
\colhead{Fluxratios} &
\colhead{} &
\colhead{Separation}  &
\colhead{Position angle\tablenotemark{a}}\\
\colhead{designation} &
\colhead{[$J$] 1.26\,\micron} &
\colhead{[$H$] 1.65\,\micron} &
\colhead{[$K^{\prime}$] 2.13\,\micron} &
\colhead{($''$)}  &
\colhead{($\arcdeg$)}
}
\startdata
HD\,17332       & \nodata         & 1.58 $\pm$ 0.05 & \nodata          & 3.66  $\pm$ 0.02  & 309.39 $\pm$ 0.02\\ 
RXJ\,0255.3+1915& 31.5 $\pm$ 0.5~\, & 24.7 $\pm$ 0.1~\, & 24    $\pm$ 5\tablenotemark{b}~\, & 1.025 $\pm$ 0.004 & 160.27 $\pm$ 0.03\\
RXJ\,0255.4+2005& 1.01 $\pm$ 0.01 & 1.00 $\pm$ 0.01 & 1.05  $\pm$ 0.05 & 0.533 $\pm$ 0.003 & 101.81 $\pm$ 0.04\\
LkH$\alpha$\,263& 1.25 $\pm$ 0.03 & 1.10 $\pm$ 0.02 & 0.89  $\pm$ 0.05 & 0.416 $\pm$ 0.003 & 232.67 $\pm$ 0.05\\
E\,02553+2018   & 1.06 $\pm$ 0.05 & 1.37 $\pm$ 0.03 & 1.7   $\pm$ 0.2  & 1.144 $\pm$ 0.005 & 163.77 $\pm$ 0.03\\
S18 A/(Ba+Bb)   & 1.25 $\pm$ 0.02 & 1.35 $\pm$ 0.01 & 1.3   $\pm$ 0.2  & 0.747 $\pm$ 0.005 & 130.34 $\pm$ 0.4~~\\
S18 Ba/Bb       & \nodata         & 1.13 $\pm$ 0.05 & \nodata          & 0.063 $\pm$ 0.004 & 56.52 $\pm$ 1.6\\
UY\,Aur         & \nodata         & 1.47 $\pm$ 0.01 & \nodata          & 0.894 $\pm$ 0.004 & 228.82 $\pm$ 0.03\\
TWA\,2          & \nodata         & 1.98 $\pm$ 0.07 & \nodata          & 0.547 $\pm$ 0.003 & ~~30.49 $\pm$ 0.04\,\\
TWA\,3          & \nodata         & 1.8  $\pm$ 0.1  & \nodata          & 1.477 $\pm$ 0.006 & 215.78 $\pm$ 0.09\\
TWA\,5\,(Aa+Ab)/B  & 120 $\pm$ 20~\, & 112 $\pm$ 13~\, & 98 $\pm$ 26\tablenotemark{b} 
                & 1.954 $\pm$ 0.008 & 359.16 $\pm$ 0.08\\
TWA\,5\,Aa/Ab   & 0.94 $\pm$ 0.05 & 1.09 $\pm$ 0.08 & 1.11  $\pm$ 0.07 & 0.054 $\pm$ 0.003 &  24.15 $\pm$ 2.8\\
\enddata

\tablenotetext{a}{Position angles are measured from north to east. Quoted errors
are relative; the systematic errors, due to array orientation uncertainties, are 
likely to be $\sim$ $1\fdg$}

\tablenotetext{b}{These relatively large errors are due to unfortunate 
placements of the secondary on some bad pixels in the array.}

\end{deluxetable}

\clearpage
\begin{deluxetable}{lrrccc}
\tabletypesize{\scriptsize}
\tablecaption{Inferred photometry of multiple system components\label{tbl-2}}
\tablewidth{0pt}
\tablecolumns{6}
\tablehead{
\colhead{Object} &
\colhead{$\alpha$(2000.0)} &
\colhead{$\delta$(2000.0)} &
\colhead{} &
\colhead{IR magnitudes\tablenotemark{a}}  &
\colhead{} \\
\colhead{designation} &
\colhead{(h m s)} &
\colhead{(\arcdeg~\arcmin~\arcsec)}  &
\colhead{[$J$]} &
\colhead{[$H$]} &
\colhead{[$K_s$]}
}
\startdata
HD\,17332     A   & 02 47 27.42 &  19 22 18.6 & \nodata           & \,6.10 $\pm$ 0.03 & \nodata \\ 
HD\,17332     B   &             &             & \nodata           & \,6.59 $\pm$ 0.04 & \nodata \\ 
RXJ\,0255.3+1915 A& 02 55 16.60 &  19 15 01.5 & \,9.36 $\pm$ 0.02 & \,9.10 $\pm$ 0.03 & ~9.02 $\pm$ 0.03 \\
RXJ\,0255.3+1915 B&             &             & 13.10 $\pm$ 0.04~ & 12.58 $\pm$ 0.03~ & 12.49 $\pm$ 0.25\, \\
RXJ\,0255.4+2005 A& 02 55 25.78 &  20 04 51.7 & 10.54 $\pm$ 0.03~ & \,9.93 $\pm$ 0.03 & ~9.72 $\pm$ 0.05 \\
RXJ\,0255.4+2005 B&             &             & 10.54 $\pm$ 0.03~ & \,9.93 $\pm$ 0.03 & ~9.77 $\pm$ 0.05 \\
LkH$\alpha$\,263 A& 02 56 08.42 &  20 03 38.6 & 11.26 $\pm$ 0.03~ & 10.55 $\pm$ 0.03~ & 10.32 $\pm$ 0.05\, \\
LkH$\alpha$\,263 B&             &             & 11.50 $\pm$ 0.04~ & 10.65 $\pm$ 0.03~ & 10.19 $\pm$ 0.05\, \\
E\,02553+2018 A   & 02 58 11.23 &  20 30 03.5 & \,9.98 $\pm$ 0.05 & \,9.09 $\pm$ 0.04 & ~8.60 $\pm$ 0.06 \\
E\,02553+2018 B   &             &             & 10.04 $\pm$ 0.05~ & \,9.43 $\pm$ 0.04 & ~9.18 $\pm$ 0.09 \\
S18 A             & 03 02 21.05 &  17 10 34.2 & 11.31 $\pm$ 0.03~ & 10.49 $\pm$ 0.03~ & 10.18 $\pm$ 0.09\, \\
S18 B(a+b)        &             &             & 11.55 $\pm$ 0.03~ & 10.82 $\pm$ 0.03~ & 10.47 $\pm$ 0.11\, \\
S18 Ba            &             &             & \nodata           & 11.51 $\pm$ 0.05~ & \nodata \\
S18 Bb            &             &             & \nodata           & 11.64 $\pm$ 0.05~ & \nodata \\
UY\,Aur A         & 04 51 47.38 &  30 47 13.4 & \nodata           & \,8.26 $\pm$ 0.07\tablenotemark{b} & \nodata \\
UY\,Aur B         &             &             & \nodata           & \,8.68 $\pm$ 0.10\tablenotemark{b} & \nodata \\
TWA\,2 A          & 11 09 13.81 & -30 01 39.8 & \nodata           & \,7.37 $\pm$ 0.05 & \nodata \\
TWA\,2 B          &             &             & \nodata           & \,8.11 $\pm$ 0.07 & \nodata \\
TWA\,3 A          &11 10 28.0~\,&-37 31 53~~\,& \nodata           & \,7.53 $\pm$ 0.05 & \nodata \\
TWA\,3 B          &             &             & \nodata           & \,8.15 $\pm$ 0.07 & \nodata \\
TWA\,5\,Aa        & 11 31 55.27 & -34 36 27.4 & \,8.46 $\pm$ 0.05 & \,7.69 $\pm$ 0.07 & ~7.45 $\pm$ 0.05 \\
TWA\,5\,Ab        &             &             & \,8.39 $\pm$ 0.05 & \,7.79 $\pm$ 0.08 & ~7.56 $\pm$ 0.06 \\
\enddata

\tablecomments{Coordinates are from Luhman (2001), Webb et al. (1999) or the
Tycho-2 catalog (H{\o}g et al. 2000). The absolute photometry has been inferred
from 2MASS measurements (the All-Sky Data Release) of the unresolved systems
together with our obtained flux ratios.
}

\tablenotetext{a}{The reported uncertainties are only the propagated errors from the 
2MASS photometry and our flux ratios. The epoch difference ($\sim$1998 for 2MASS and 2000
for our data) introduces an additional, but unknown, error due to the variability of
the stars.}

\tablenotetext{b}{For the primary UY\,Aur~A we adopt the H magnitude of Close et al. 
(1998), and then assume that the observed change of flux ratio is entirely due to the 
companion getting brighter. See also section \ref{UYAur}.
}
\end{deluxetable}

\clearpage
\begin{deluxetable}{lllc}
\tabletypesize{\scriptsize}
\tablecaption{Stars without detected close ($\lesssim$1\farcs6) companions\label{tbl-3}}
\tablewidth{0pt}
\tablecolumns{6}
\tablehead{
\colhead{Object} &
\colhead{$\alpha$(2000.0)} &
\colhead{$\delta$(2000.0)} & 
\colhead{Observed}
\\
\colhead{designation} &
\colhead{(h m s)} &
\colhead{(\arcdeg~\arcmin~\arcsec)}  &
\colhead{band}
}
\startdata 
LkH$\alpha$\,262 & 02 56 07.99 & ~20 03 24.3 & $H$,$K^{\prime}$ \\
LkH$\alpha$\,264 & 02 56 37.56 & ~20 05 37.1 & $H$,$K^{\prime}$ \\
RX\,J0258.3+1947 & 02 58 16.09 & ~19 47 19.6 & $J$,$H$ \\
RX\,J0306.1+1921 & 03 06 33.1  & ~19 21 52   & $J$,$H$,$K^{\prime}$ \\
TWA\,6           & 10 18 28.8  & -31 50 02   & $H$ \\
TWA\,7           & 10 42 30.3  & -33 40 17   & $H$ \\
TWA\,1           & 11 01 51.9  & -34 42 17   & $H$ \\
TWA\,8~B         & 11 32 41.4  & -26 52 08   & $H$ \\
TWA\,8~A         & 11 32 41.5  & -26 51 55   & $H$ \\
TWA\,9~B         & 11 48 23.6  & -37 28 49   & $H$ \\
TWA\,9~A         & 11 48 24.2  & -37 28 49   & $H$ \\
TWA\,10          & 12 35 04.3  & -41 36 39   & $H$ \\
TWA\,11~B        & 12 36 00.8  & -39 52 15   & $H$ \\
TWA\,11~A        & 12 36 01.3  & -39 52 09   & $H$ \\
\enddata

\tablecomments{Coordinates are from Luhman (2001), Hearty et al. (2000), Webb et al. 
(1999) or the Tycho-2 catalog (H{\o}g et al. 2000). Our observations are sensitive
to separations between 0\farcs042 (diffraction limit at $H$)
and 1\farcs6 (smallest field of view radius, but in some directions out to 3\farcs4). The
contrast sensitivity as a function of separation is depicted in Fig.~\ref{fig-1}.
}
\end{deluxetable}

\begin{references}
\reference{} Bate, M. R., Bonnell, I. A. \& Bromm, V. 2002, MNRAS, 336, 705
\reference{} Brandeker, A., Liseau, R., Artumowicz, P., \& Jayawardhana, R. 2001, 
     \apjl, 561, L199
\reference{} Chauvin, G., M{e}\'nard, F., Fusco, T., Lagrange, A.-M., Beuzit, J.-L., 
     Mouillet, D., \& Augerau, J.-C. 2002, \aap, 394, 949
\reference{} Christou, J., Bonaccini, D., Ageorges, N.,
     \& Marchis, F. 1999, ESO Messenger, 97, 14
\reference{} Close, L. M., et al. 1998, \apj, 499, 883
\reference{} Duch\^{e}ne, G. 1999, \aap, 341, 547
\reference{} Duch\^{e}ne, G., Ghez, A. M., \& McCabe, C. 2002, \apj, 568, 771
\reference{} Dutrey, A., Guilloteau, S., Duvert, G., Prato, L.,
     Simon, M., Schuster, K., \& Menard, F. 1996, \aap, 309, 493
\reference{} Duvert, G., Dutrey, A., Guilloteau, S., Menard, F.,
     Schuster, K., Prato, L., \& Simon, M. 1998, \aap, 332, 867
\reference{} Ghez, A. M., Neugebauer, G.,
     \& Matthews, K. 1993, \aj, 106, 2005
\reference{} Hearty, T., Fern\'{a}ndez, M., Alcal\'{a}, J. M.,
     Covino, E., \& Neuh\"{a}user, R. 2000, \aap, 357, 681
\reference{} Herbst, T. M., Koresko, C. D.,
     \& Leinert, C. 1995, \apjl, 444, L93
\reference{} H{\o}g, E., et al. 2000, \aap, 355, L27
\reference{} Itoh, Y., et al. 2003, \apj, 586, L141
\reference{} Hogerheijde, M., Johnstone, D., 
        Matsuyama, I., Jayawardhana, R., \& Muzerolle, J. 2003, \apj, submitted 
\reference{} Jayawardhana, R., Hartmann, L., Fazio, G., 
        Fisher, R.S., Telesco, C.M., \& Pi\~na, R.K. 1999a, \apj, 520, L41
\reference{} Jayawardhana, R., Hartmann, L., Fazio, G., 
        Fisher, R.S., Telesco, C.M., \& Pi\~na, R.K. 1999b, \apj, 521, L129
\reference{} Jayawardhana, R. \& Greene, T.P. (ed.) 2001, {\it Young Stars Near Earth:
      Progress and Prospects}, ASP Conf. Ser. 244 (San Fransisco: ASP)
\reference{} Jayawardhana, R., Wolk, S.J., Barrado y 
        Navascu\'es, D., Telesco, C.M., \& Hearty, T.J. 2001, \apj, 550, L197
\reference{} Jayawardhana, R., Luhman, K. L., D'Alessio, P.,
     \& Stauffer, J. R. 2002, \apjl, 571, L51
\reference{} Joy, A. H., \& van Biesbroeck, G. 1944, \pasp, 56, 123
\reference{} Kastner, J. H., Zuckerman, B., Weintraub, D. A.,
     \& Forveille, T. 1997, Science, 277, 67
\reference{} Leinert, C., Zinnecker, H., Weitzel, N., Christou, J.,
      Ridgway, S. T., Jameson, R., Haas, M., \& Lenzen, R. 1993, \aap, 278, 129
\reference{} Lowrance, P. J., et al. 1999, \apjl, 512, L69
\reference{} Luhman, K. L. 2001, \apj, 560, 287
\reference{} Macintosh, B., et al. 2001, in ASP Conf. Ser. 244, {\it Young Stars Near Earth: 
      Progress and Prospects}, ed. R. Jayawardhana \& T. P. Greene (San Fransisco: ASP), 309
\reference{} Mathieu, R. D. 1994, \araa, 32, 465
\reference{} Mathis, J. S. 1990, \araa, 28, 37
\reference{} Neuh\"{a}user, R., Guenther, E. W., Petr, M. G.,
     Brandner, W., Hu\'{e}lamo, N., \& Alves, J. 2000, \aap, 360, L39
\reference{} Palla, F., \& Stahler, S. W. 2001, \apj, 553, 299
\reference{} Perryman, M. A. C., et al. 1997, \aap, 323, L49
\reference{} Steffen, A., et al. 2001, \apj, 122, 997
\reference{} Torres, G., Guenther, E. W., Marschall, L. A.,
     Neuh\"{a}user, R., Latham, D. W., \& Stefanik, R. P. 2003, \aj, 125, 825
\reference{} Webb, R. A., Zuckerman, B., Platais, I., Patience, J.,
     White, R. J., Schwartz, M. J., \& McCarthy, C. 1999, \apjl, 512, L63
\reference{} Weintraub, D. A., Saumon, D., Kastner, J. H.,
     \& Forveille, T. 2000, \apj, 530, 867
\reference{} White, R. J., \& Ghez, A. M. 2001, \apj , 556, 265
\reference{} Wizinowich, P., et al. 2000, \pasp, 112, 315
\reference{} Zuckerman, B., Webb, R. A., Schwartz, M., \& Becklin, E. E. 2001, \apjl, 549, L233
\end{references}
\end{document}